\global\def\draftcontrol{0}

   \def\versionno{ ktqnm}

\catcode`\@=11

\expandafter\ifx\csname draftcontrol\endcsname\relax\global\def\draftcontrol{0}
\fi

{\count255=\time\divide\count255 by 60
\xdef\hourmin{\number\count255}
\multiply\count255 by-60\advance\count255 by\time
\xdef\hourmin{\hourmin:\ifnum\count255<10 0\fi\the\count255}}
\def\draftdate{\number\month/\number\day/\number\year\ \ \ \hourmin }

\newcommand\makepapertitle{\par
  \begingroup
    \renewcommand\thefootnote{\@fnsymbol\c@footnote}%
    \def\@makefnmark{\rlap{\@textsuperscript{\normalfont\@thefnmark}}}%
    \long\def\@makefntext##1{\parindent 1em\noindent
            \hb@xt@1.8em{%
                \hss\@textsuperscript{\normalfont\@thefnmark}}##1}%
     \newpage
     \global\@topnum\z@   
     \@makepapertitle
     \thispagestyle{empty}\@thanks
  \endgroup
  \setcounter{footnote}{0}%
  \global\let\thanks\relax
  \global\let\makepapertitle\relax
  \global\let\@makepapertitle\relax
  \global\let\@thanks\@empty
  \global\let\@author\@empty
  \global\let\@date\@empty
  \global\let\@title\@empty
  \global\let\title\relax
  \global\let\author\relax
  \global\let\date\relax
  \global\let\and\relax
  \def\version{\let\version\@version\@gobble}
}
\def\@makepapertitle{%
  \newpage
   \ifnum\draftcontrol=1 {}
   \version\versionno
   \vskip 3em%
   \else
   \hfill\hbox to 3cm {\parbox{4cm}{\@pubnum}\hss}%
   \vskip 3em%
   \fi
   \begin{center}%
   \let \footnote \thanks
     {\LARGE {\@title}}%
     \vskip 1.5em%
     {\normalsize
       \lineskip .5em%
       \begin{tabular}[t]{c}%
         \@author
       \end{tabular}\par}%
     \vskip 1.5em%
     {\@bstract}%
     \end{center}%
     \vskip 1.5em
     \@date%
   \par
}

\gdef\@pubnum{}
\def\pubnum#1{%
  \gdef\@pubnum{#1}}

\gdef\@bstract{}
\def\Abstract#1{%
  \gdef\@bstract{%
   \parbox{\textwidth-0pc}{%
   \centerline{\bf Abstract}\penalty1000%
\kern.2cm%
\noindent
\renewcommand\baselinestretch{1.0}%
{#1}}}
}

\def\ps@paper{\let\@mkboth\@gobbletwo%
     \ifnum\draftcontrol=1
    \def\@oddfoot{\hbox to \textwidth{\tiny \versionno \hfil\tiny\draftdate}%
    \hskip -\textwidth \hbox to \textwidth{\hfil\rm\thepage\hfil}}%
     \else\def\@oddfoot{\hbox to \textwidth{\hfil\rm\thepage\hfil}}
     \fi
     \let\@evenfoot\@oddfoot
}

\def\body{\clearpage
          \pagestyle{paper}
    }

\def\@version#1{\ifnum\draftcontrol=1
\typeout{}\typeout{#1}\typeout{}
\vskip3mm\centerline{\hbox{\fbox{\normalsize{\tt DRAFT -- #1 -- }
                   {\draftdate}}}}\vskip3mm
\fi}
\let\version\@version
\long\def\eqlabel#1{\ifnum\draftcontrol=1
                    \tag@false  
                    \tag*{(\theequation) \hbox to -0.2cm{\hspace{0cm}\small{#1}\hss}}
                    \refstepcounter{equation}
                    \edef\@currentlabel{\theequation}
                    \ltx@label{#1}          
                    \else
                    \label{#1}
                    \fi
                    }
\let\st@bibitem\@bibitem
\let\st@lbibitem\@lbibitem
\ifnum\draftcontrol=1
  \def\@bibitem#1{%
    \st@bibitem{#1}\a@@label{#1}\ignorespaces}
  \def\@lbibitem[#1]#2{%
    \st@lbibitem[#1]{#2}\a@@label{#2}\ignorespaces}
  \def\a@@label#1{%
    \gdef\a@lab{\smash{\normalfont\small#1}}
    \ifvmode
      \if@inlabel
        \global\setbox\@labels\hbox{%
          \llap{\a@lab\let\a@lab\relax
                \kern\@totalleftmargin\kern\marginparsep}%
          \box\@labels}%
      \fi
    \fi}
\fi

\documentclass[12pt,letterpaper]{article}

\usepackage{amsmath,amssymb,array,calc,epsfig,rotating,bm}
\usepackage[sort]{cite}
\usepackage{graphicx}
\usepackage{psfrag,verbatim}
\usepackage{xcolor}

\ifnum\draftcontrol=1
\fi

\renewcommand\baselinestretch{1.25}
\renewcommand\section{\@startsection {section}{1}{\z@}%
                                   {-3.5ex \@plus -1ex \@minus -.2ex}%
                                   {2.3ex \@plus.2ex}%
                                   {\normalfont\large\bfseries}}
\renewcommand\subsection{\@startsection{subsection}{2}{\z@}%
                                   {-3.25ex\@plus -1ex \@minus -.2ex}%
                                   {1.5ex \@plus .2ex}%
                                   {\normalfont\normalsize\bfseries}}
\renewcommand\subsubsection{\@startsection{subsubsection}{3}{\z@}%
                                   {-3.25ex\@plus -1ex \@minus -.2ex}%
                                   {1.5ex \@plus .2ex}%
                                   {\normalfont\normalsize\it}}
\renewcommand\paragraph{\@startsection{paragraph}{4}{\z@}%
                                   {-3.25ex\@plus -1ex \@minus -.2ex}%
                                   {1.5ex \@plus .2ex}%
                                   {\normalfont\normalsize\bf}}

\numberwithin{equation}{section}

\def\revise#1       {\raisebox{-0em}{\rule{3pt}{1em}}%
                     \marginpar{\raisebox{.5em}{\vrule width3pt\
                     \vrule width0pt height 0pt depth0.5em
                     \hbox to 0cm{\hspace{0cm}{%
                     \parbox[t]{4em}{\raggedright\footnotesize{#1}}}\hss}}}}

\newcommand\nxt[1]  {\\\fnxt#1}
\newcommand{\ie}{{\it i.e.,}\ }

\def\calb         {{\cal B}}

\def\cale         {{\cal E}}

\def\calj         {{\cal J}}
\def\calk         {{\cal K}}
\def\call         {{\cal L}}
\def\calm         {{\cal M}}

\def\calo         {{\cal O}}

\def\zet          {{\mathbb Z}}

\def\qnm         {{\hat{q}}}

\def\sqr#1#2{{\vcenter{\vbox{\hrule height.#2pt
 \hbox{\vrule width.#2pt height#1pt \kern#1pt
 \vrule width.#2pt}\hrule height.#2pt}}}}
\def\square{%
  \mathop{\mathchoice{\sqr{12}{15}}{\sqr{9}{12}}{\sqr{6.3}{9}}{\sqr{4.5}{9}}}}

\newcommand{\kk}{\mathfrak{q}}

\def\dd{\delta}

\def\aa1{\phi}
\def\cc1{\psi}

\def\csb{{\chi\rm{SB}}}

\catcode`\@=12

\begin{document}


\title{\bf On branches of the KS black hole}

\date{May 2, 2020}

\author{
Alex Buchel \\
\it Department of Applied Mathematics\\
\it Department of Physics and Astronomy\\ 
\it University of Western Ontario\\
\it London, Ontario N6A 5B7, Canada\\
\it Perimeter Institute for Theoretical Physics\\
\it Waterloo, Ontario N2J 2W9, Canada\\[0.4cm]
}

\Abstract{The Klebanov-Strassler black hole is a holographic dual to ${\cal N}=1$
supersymmetric $SU(N)\times SU(N+M)$ cascading gauge theory plasma
with spontaneously broken chiral symmetry. The chiral symmetry breaking
sector of the cascading gauge theory contains two dimension-3
operators and a single dimension-7 operator. The black hole solution
constructed in \cite{Buchel:2018bzp} represents the end point of the
instability triggered by the condensation of one of the dimension-3
operators. We study here all three branches of the quasinormal modes
of the chiral symmetry breaking sector --- there are no additional
instabilities beyond the one identified in \cite{Buchel:2010wp}.
Thus, the Klebanov-Strassler black hole solution of \cite{Buchel:2018bzp}
is the only one with homogeneous and isotropic horizon,
perturbatively connected to the chirally symmetric Klebanov-Tseytlin
black hole \cite{Aharony:2007vg}.
}

\makepapertitle

\body

\version\versionno
\tableofcontents

\section{Introduction and summary}\label{intro}

The Klebanov-Strassler (KS) black hole \cite{Buchel:2018bzp} is a holographic dual
to a cascading gauge theory\footnote{See \cite{Herzog:2001xk} for the details of the
holographic correspondence.} \cite{Klebanov:2000hb}
plasma in the deconfined homogeneous and isotropic state with spontaneously broken chiral symmetry. 
It is a ``branch'' of the Klebanov-Tseytlin (KT) black hole \cite{Buchel:2000ch,
Buchel:2001gw,Gubser:2001ri,Aharony:2005zr,Aharony:2007vg} --- a holographic dual to a
deconfined thermal equilibrium state of the cascading gauge theory with unbroken
chiral symmetry --- in that it is associated with the perturbative spontaneous
chiral symmetry breaking ($\csb$). In other word, there is a critical temperature
\cite{Buchel:2010wp}
\begin{equation}
T_\csb=0.54195\ \Lambda\,,
\eqlabel{tcrit}
\end{equation}
with $\Lambda$ being the strong coupling scale of the cascading gauge theory
\cite{Aharony:2007vg,Bena:2019sxm}, below which the $\csb$ fluctuations become
perturbatively unstable. The endpoint of condensation of these fluctuations
is the KS black hole \cite{Buchel:2018bzp}. The constructed KS black hole 
has some unusual features, when interpreted as a holographic dual of a thermal
equilibrium state of the renormalizable\footnote{Holographic renormalization
of the cascading gauge theory was established in \cite{Aharony:2005zr}.} four-dimensional
gauge theory:
\begin{itemize}
\item{(i)} To begin, the KS black hole is actually neither thermodynamically
nor dynamically stable --- it has a negative specific heat, and correspondingly  \cite{Buchel:2005nt}
an imaginary speed of the sound waves. The latter implies that the small inhomogeneities
in the energy density and the pressure in the chiral
symmetry broken phase of the cascading plasma amplify, destroying the homogeneity and isotropy
of the corresponding thermal state.
\item{(ii)} The phase transitions in the cascading gauge theory plasma look very different in
canonical and microcanonical ensembles. From the canonical ensemble perspective,
the KS black hole is an {\it exotic}
object\footnote{First examples of such objects in holography were
reported in \cite{Buchel:2009ge}.} --- it exists only for $T\ge T_\csb$ and has a higher free energy
density as compared to that of the KT black at the corresponding temperature. It is thus
irrelevant in the canonical ensemble.
On the other hand, in the microcanonical ensemble, it is a dominant (more entropic) configuration
below some critical energy density $\cale_{\csb}$ (associated with $T_{\csb}$) --- in a
constraint (to spatial homogeneity and isotropy) dynamical evolution it is the end-point
of the relaxation associated with the $\csb$.
\item{(iii)} The $\csb$ instability occurs at a lower temperatures/lower energy densities than the
confinement/deconfinement phase transition 
in the cascading gauge theory \cite{Aharony:2007vg},
\begin{equation}
T_c= 0.6141111\ \Lambda\ >\ T_{\csb}\,.
\eqlabel{tc}
\end{equation}
Here the (canonical ensemble) transition proceeds from the 
deconfined chirally symmetric phase to a confined phase with spontaneous $\csb$.
Still, $T_{\csb}$ is above the temperature of the
thermodynamic/dynamic instability in the KT black hole plasma \cite{Buchel:2009bh},
\begin{equation}
T_\csb\ >\ T_u=0.537286\ \Lambda\,.
\eqlabel{tu}
\end{equation}
\end{itemize}

While unusual, (i)-(iii) can be incorporated into a consistent
physics story.
\nxt (i) and (ii) suggests  that  the end point of the $\csb$ in the  cascading plasma
can simply be an inhomogeneous state; it would proceed in two steps:
spontaneous symmetry breaking as in \cite{Buchel:2010wp}, followed by the relaxation to
a spatially inhomogeneous equilibrium state (so far unknown). 
There are plenty examples of such phenomena (albeit in the presence of the chemical
potential/charge density), see  \cite{Nickel:2009wj} for examples in the Nambu-Jona-Lasino-type models,
and \cite{Donos:2011bh} for a holographic model. 
\nxt (iii) suggests that the confinement and the chiral symmetry breaking can be two separate transitions,
see \cite{Gross:1991pk} for an example in a phenomenological QCD model and \cite{Aharony:2006da}
for a holographic example. Despite \eqref{tc}, the cascading gauge theory $\csb$ phase transition
can still be important as it is the second-order (thus being a perturbative one) phase transition,
while the confinement transition of  \cite{Aharony:2007vg}
is of the first-order, proceeding through the bubble nucleation, which is strongly suppressed
for a large number of colors. 

Alternatively, it is possible that there are other KS black holes, beyond the construction of
\cite{Buchel:2018bzp}, with the more standard thermodynamics and the
pattern of the phase transitions. This is the question we would like
to address in this paper. It is conceivable that there exist KS black holes separated from the
KT black hole by the first-order phase-transition --- we can not add anything new here.
Instead, we focus on potentially additional ``branches'' of the KS black holes, connected to
the KT black hole ``trunk'' via instabilities in the chiral symmetry breaking sector. 
The reason why one might expect additional branches is related to the richness of the chiral
symmetry breaking sector: as explain in \cite{Buchel:2010wp}, this sector involves
linearized gravitational fluctuations dual to a pair of dimension-3 operators and a
single dimension-7 operator. Thus, from the holographic bulk perspective one expects three distinct
branches of the quasinormal modes (QNMs) of the KT black hole. Only a single branch
(referred to as $\calb_{3u}$ here) was identified in  \cite{Buchel:2010wp}.

In what follows, we identify all the three branch of the $\csb$
QNMs: $\calb_{3u}$, $\calb_{3s}$ and $\calb_{7}$.
We find that the former two branches in the limit $T\gg \Lambda$ combine into a single branch representing
$\Delta=3$ QNMs of the $AdS_5$-Schwarzschild black hole \cite{Nunez:2003eq}.  At high temperatures,
the degeneracy between the two branches is broken by $\calo(1/\sqrt{\ln\frac{T}{\Lambda}})$ effects.
Following these branches to low temperatures, we recover the instability on the $\calb_{3u}$ branch
at $T_\csb$ \eqref{tcrit}, originally found in \cite{Buchel:2010wp}. The other branch,
$\calb_{3s}$, corresponds to the QNMs that remain stable for $T>T_{\csb}$. The $\calb_7$ branch
at high temperatures represents  $\Delta=7$ QNMs of the $AdS_5$-Schwarzschild black
hole \cite{Nunez:2003eq}. The dispersion relation of its QNMs differ from that of
the conformal modes  by $\calo(1/\ln\frac{T}{\Lambda})$ effects. Similar to the  $\calb_{3s}$
branch, the $\calb_{7}$ branch corresponds to the QNMs that remain stable for  $T>T_{\csb}$.
Thus, we conclude that the $\csb$ instability  of \cite{Buchel:2010wp} is the dominant
perturbative instability of the Klebanov-Tseytlin black hole
for $T\ge T_{\csb}$. Of course, this is the instability of the lowest QNM on the $\calb_{3u}$ branch
--- as one further reduces the temperature below $T_{\csb}$, one expects developing
instabilities of the excited QNMs on $\calb_{3u}$, as well as on the other two branches of the $\csb$
sector. 

In the rest of this section we present the numerical results for the spectrum of the QNMs on the three
branches of the $\csb$ sector. Specifically, we analyze $\kk^2\equiv k^2/(2\pi T)^2$ of the QNMs
at zero frequency, $\omega=0$, as one varies the Hawking temperature of the KT black hole.
A QNM becomes unstable once $\kk^2>0$. The technical details relevant to the reported results
are collected in section \ref{eoms} and the appendices \ref{app7} and \ref{app3}.
In section \ref{sussec} we present an independent argument for the absence of the 
KS black holes (in addition to the one constructed in \cite{Buchel:2018bzp}),
perturbatively related to the KT black hole.

\begin{figure}[t]
\begin{center}
\psfrag{q}[cc][][1][0]{$\kk^2$}
\psfrag{b}[cc][][1][0]{$\ln b$}
\includegraphics[width=3in]{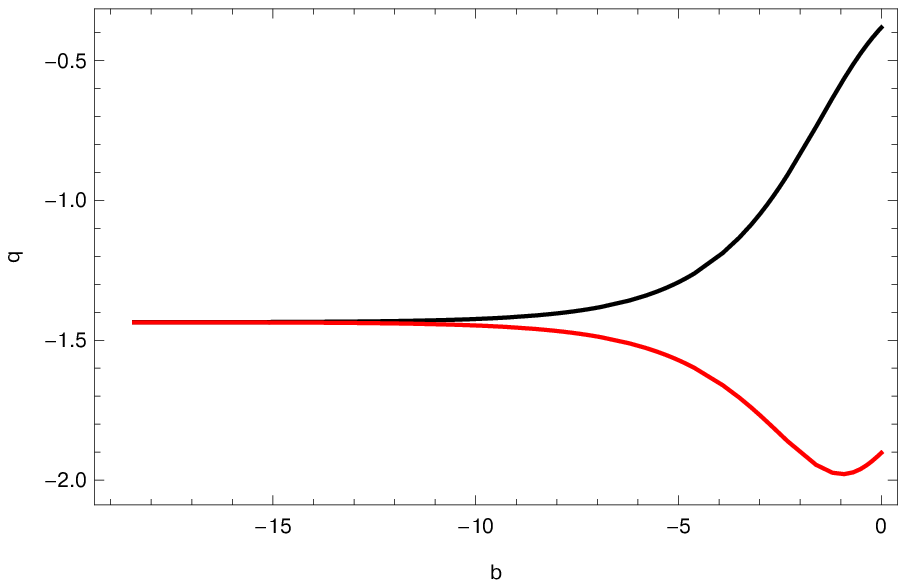}\,
\includegraphics[width=3in]{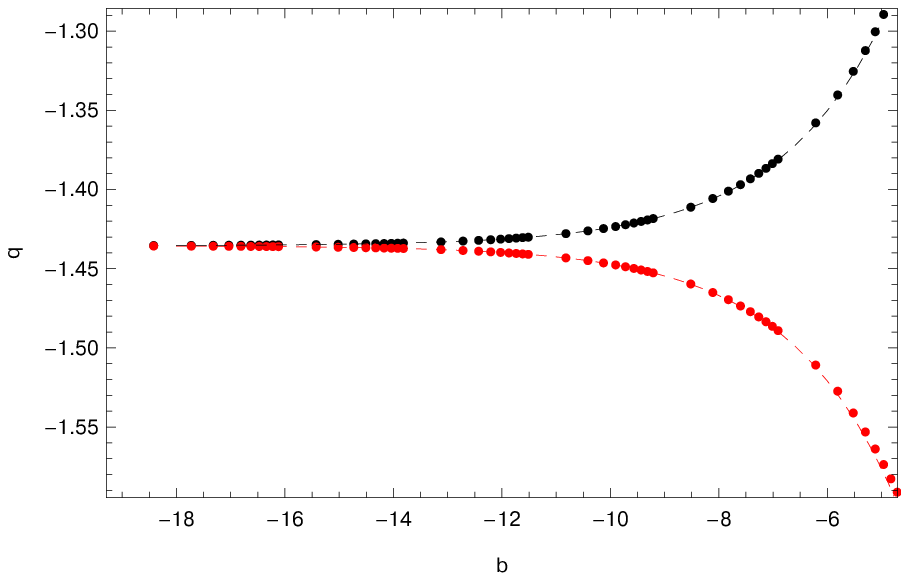}
\end{center}
  \caption{Functional dependence of $\kk^2\equiv k^2/(2\pi T)^2$ of the quasinormal modes
   of the $\calb_{3u}$ branch (black) and  of the quasinormal modes of the $\calb_{3s}$ branch  (red)
   at $\omega=0$ in the
  high-temperature/small $b$ (see \eqref{btl}) regime of the KT black hole.
These quasinormal modes are dual to the fluctuations of the $\csb$ gaugino condensate operators in
the symmetric phase of the cascading gauge theory plasma. 
The dashed curves in the right panel
  show the leading order high-temperature corrections to the conformal value of $\kk^2$ ---
  a temperature independent result for the quasinormal mode dual to the dimension-3 operator in
  the holographic $CFT_4$. The QNMs are stable since $\kk^2<0$, in the temperature regime reported.  
} \label{figure33ahigh}
\end{figure}

\begin{figure}[t]
\begin{center}
\psfrag{q}[cc][][1][0]{$\kk^2$}
\psfrag{b}[cc][][1][0]{$\ln b$}
\includegraphics[width=3in]{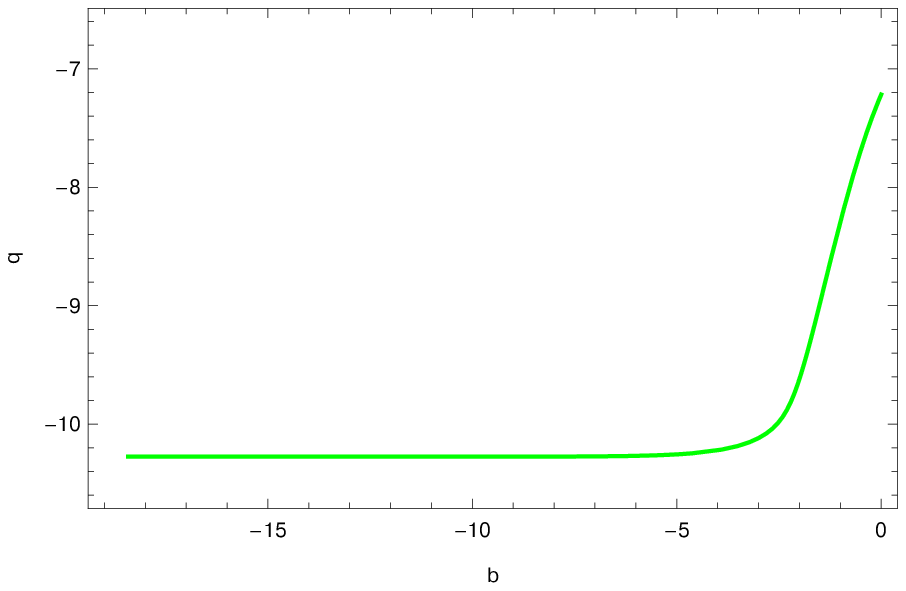}\,
\includegraphics[width=3in]{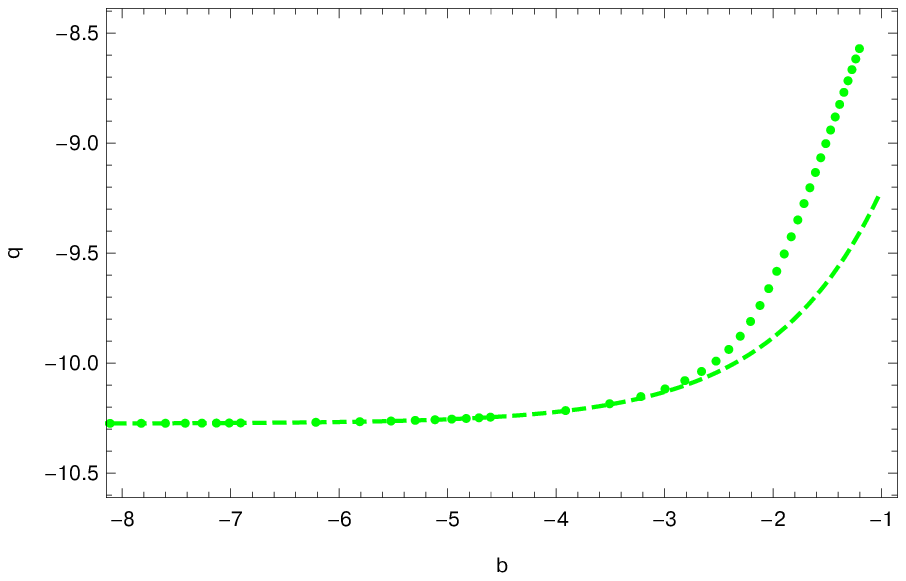}
\end{center}
  \caption{Functional dependence of $\kk^2\equiv k^2/(2\pi T)^2$  of the quasinormal modes of the $\calb_7$ branch
  at $\omega=0$ in the
  high-temperature/small $b$ (see \eqref{btl}) regime of the KT black hole.
These quasinormal modes are dual to the fluctuations of the $\csb$ dimension-7 operator in
the symmetric phase of the cascading gauge theory plasma. 
The dashed curve in the right panel
  shows the leading order high-temperature correction to the conformal value of $\kk^2$ ---
  a temperature independent result for the quasinormal mode dual to the dimension-7 operator in
  the holographic $CFT_4$. The QNMs are stable since $\kk^2<0$, in the temperature regime reported.  
} \label{figure7high}
\end{figure}

Figs.~\ref{figure33ahigh} and \ref{figure7high} present the results for the $\kk^2$ (dimensionless momentum,
see \eqref{defqq}) of the lowest QNMs for the three branches $\calb_{3u}$ (black curves),
$\calb_{3s}$ (red curves) and $\calb_7$ (green curves) at zero frequency $\omega=0$ and high temperature,
see section \ref{hight}. In the limit $b\to 0$,
\begin{equation}
\ln b \approx -\ln \left(2\ln \frac{T}{\Lambda}\right)\,.
\eqlabel{btl}
\end{equation}
The fact that $\kk^2<0$ for all the QNMs implies that the chiral symmetry breaking fluctuations
in the KT black hole are massive, and exponentially decay both in time and space when excited.  
The dashed curves in the right panels indicate the values of $\kk^2$ for the corresponding QNM branch
in the conformal limit, along with the leading order corrections: $\calo(b)$ of \eqref{br7q2} for the
$\calb_7$ branch, and $\calo(\sqrt{b})$  for the $\calb_{3u}$ (plus sign in \eqref{br3q2})
and $\calb_{3s}$ (minus sign in \eqref{br3q2}).

\begin{figure}[t]
\begin{center}
\psfrag{q}[cc][][1][0]{$\kk^2$}
\psfrag{t}[cc][][1][0]{{$\frac{T}{\Lambda}$}}
\includegraphics[width=3in]{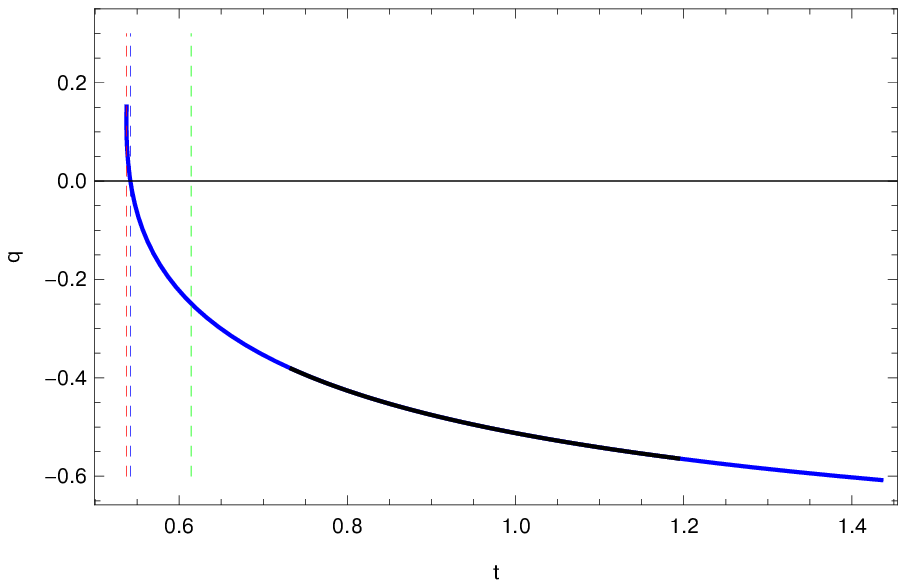}\,
\includegraphics[width=3in]{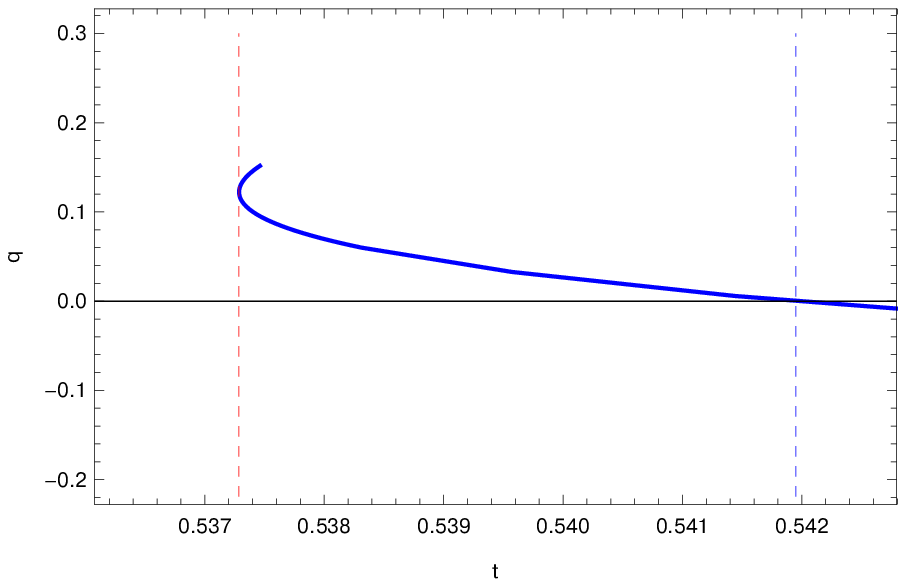}
\end{center}
  \caption{$\kk^2$ of the lowest QNM on the $\calb_{3u}$ branch as
  a function of $\frac{T}{\Lambda}$. $\kk^2>0$ for $T<T_{\csb}$ (the vertical dashed blue lines),
  signaling the perturbative instability. The blue curve represents the results in
the  computational SchemeA, and the black curve represents the results in the computational
SchemeB (the low-temperature extension of the data presented in fig.~\ref{figure33ahigh}.
The vertical dashed red lines indicate the terminal temperature of the KT black hole, $T_u$
\eqref{tu}. The vertical green dashed line indicates the temperature of the first-order
confinement phase transition for the KT black hole, $T_c$ \eqref{tc}. 
} \label{figure3}
\end{figure}

In fig.~\ref{figure3} we follow the QNM branch $\calb_{3u}$ to low temperatures.
The results reported reproduce the analysis presented in \cite{Buchel:2010wp}.
We perform analysis in two computation schemes (see section \ref{lowt}): SchemeA (the solid blue curve)
and SchemeB (the solid black curves). The solid black curve data is the low temperature extension of the
data presented in  fig.~\ref{figure33ahigh} (also the solid black curve in the left panel
and the black dots in the right panel). 
There is an excellent agreement between the two computational schemes in the
overlapping regime (overlapping black and blue curves in the left panel)
--- the fractional difference in the results is at the level  $\propto 10^{-6}$.
The solid blue curve (in both panels) crosses $\kk^2=0$ at the location
highlighted by the vertical dashed blue line, indicating the instability of
the lowest QNM on the $\calb_{3u}$ branch. This occurs precisely at the $\csb$
phase transition \eqref{tcrit}, as found in \cite{Buchel:2010wp}. The vertical dashed
red lines highlight the temperature $T_u$ \eqref{tu}, and the vertical green dashed line
highlights the temperature $T_c$ of the confinement phase transition, see \eqref{tc}.

\begin{figure}[t]
\begin{center}
\psfrag{q}[cc][][1][0]{$\kk^2$}
\psfrag{t}[cc][][1][0]{{$\frac{T}{\Lambda}$}}
\includegraphics[width=3in]{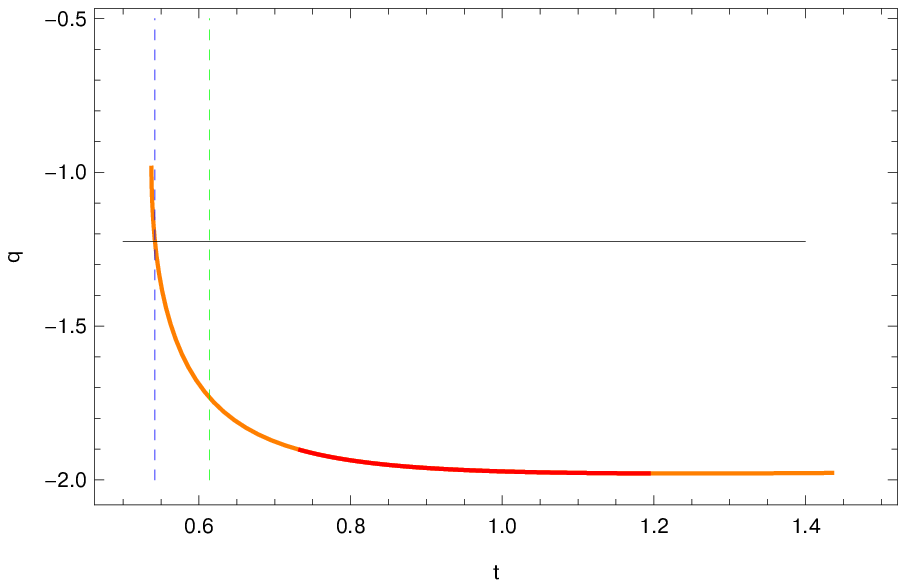}\,
\includegraphics[width=3in]{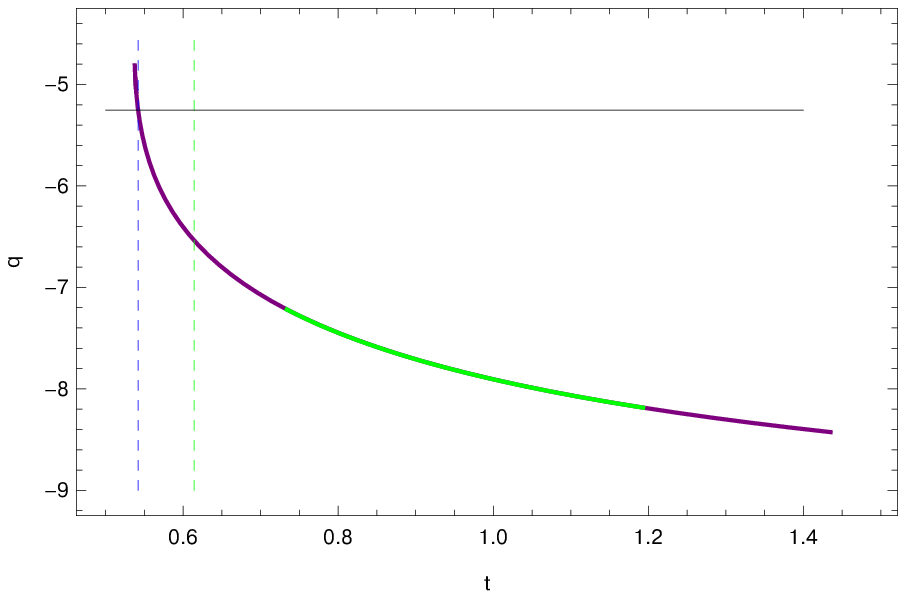}
\end{center}
  \caption{$\kk^2$ of the lowest QNMs on the $\calb_{3s}$ branch (the left panel)
  and on the $\calb_{7}$ branch (the right panel) as
  a function of $\frac{T}{\Lambda}$. $\kk^2<0$ for $T>T_{\csb}$ (the vertical dashed blue lines),
  indicating that these QNMs are stable in the temperature regime indicated.
  The orange/magenta curves represent the results in
the  computational SchemeA, and the red/green curves represent the results in the computational
SchemeB (the low-temperature extensions of the data presented in figs.~\ref{figure33ahigh} and
\ref{figure7high}.
The vertical green dashed lines indicate the temperature of the first-order
confinement phase transition for the KT black hole, $T_c$ \eqref{tc}. The horizontal
black lines denote $\kk^2(T_{\csb}/\Lambda)$ for the lowest QNM on the corresponding branch.
} \label{figure3a7}
\end{figure}

In fig.~\ref{figure3a7} we follow the QNM branches $\calb_{3s}$ (the left panel)
and $\calb_7$  (the right panel) to low temperatures.
The results reported are new. There are no instabilities on either branch for
$T> T_{\csb}$ (the vertical dashed blue lines).
The horizontal black lines indicate $\kk^2(T_\csb/\Lambda)$ for the corresponding branch. 
The vertical green dashed lines
highlight $T_c/\Lambda$. The solid orange and magenta curves are computed in SchemeA,
while the solid red and green curves are computed in SchemeB. As for the branch $\calb_{3u}$,
 the fractional differences in the results from the different schemes is at the level  $\propto 10^{-6}$.

\section{Effective action, equations of motion and the boundary asymptotes for
$\csb$ QNMs of the KT black hole}\label{eoms}

The five-dimensional effective action (KS) describing $SU(2)\times SU(2)\times \zet_2$ states of the
cascading gauge theory has been derived in \cite{Buchel:2010wp}. This effective action
contains as an on-shell solution the Klebanov-Strassler black hole
constructed in \cite{Buchel:2018bzp}.  The KS effective action allows for a consistent truncation
to a $U(1)$ chirally symmetric sector, reproducing the Klebanov-Tseytlin (KT)
effective action \cite{Aharony:2005zr}. The KT effective action contains as an on-shell solution the
KT black hole, constructed\footnote{The KT black hole at high temperatures (near conformal limit)
was discussed in \cite{Buchel:2000ch,Buchel:2001gw,Gubser:2001ri,Buchel:2009bh}.}
in \cite{Aharony:2007vg}. The effective action for the chiral symmetry breaking sector,
\ie for the $U(1)$ chiral symmetry breaking linearized fluctuations on top of the
KT states of the cascading gauge theory, was derived in \cite{Buchel:2010wp}.
 Only one branch (out of the total three branches)
of the QNMs was identified and analyzed in \cite{Buchel:2010wp}. The QNM instability
found in \cite{Buchel:2010wp} identified a bifurcation point for a branch of
the Klebanov-Strassler black holes, perturbatively connected to the KT black hole.
In this section we review the relevant effective actions, the equations of motion,
and the boundary asymptotes for
the $\csb$ QNMs of the KT black hole. Next, we proceed to a careful analysis of the QNMs in the
conformal (or high-temperature) limit, $P^2\to 0$. Although it is straightforward to identify
the three branches of the QNMs (two for the operators of $\Delta=3$ and a single one for the
operator of $\Delta=7$) in the strict $P^2=0$ (conformal) limit, constructing perturbative in $P^2$
expansions for $\Delta=3$ branches is rather subtle: while the KT BH is   
a series expansion\footnote{See \cite{Buchel:2009bh} for the KT black hole construction up to
$\calo(P^8)$ inclusive.} in $P^2$  of the $AdS_5$ Schwarzschild black hole,
we find here that the pair of $\Delta=3$ QNM branches realize a series
expansion\footnote{The same phenomenon was observed also for the normal modes of the
$\csb$ sector of the cascading gauge theory vacuum on $S^3$ in \cite{Buchel:2011cc}.}
in $\sqrt{P^2}$ --- the appearance of the square root branch point,
\ie $\pm\sqrt{P^2}$, naturally leads to breaking of the spectral degeneracy in the two
$\Delta=3$ QNM branches at finite $P^2$. Once the QNM branches are constructed
in the high-temperature (conformal) limit, it is straightforward to follow them to low-temperatures.
These results are reported in section \ref{intro}: see figs.~\ref{figure33ahigh}-\ref{figure3a7}.

\subsection{The effective actions}

The effective gravitational action describing $SU(2)\times SU(2)\times \zet_2$ states of the cascading gauge theory
is given by \cite{Buchel:2010wp}
\begin{equation}
\begin{split}
S_5=& \frac{108}{16\pi G_5} \int_{\calm_5} {\rm vol}_{\calm_5}\ \Omega_1 \Omega_2^2\Omega_3^2\ 
\biggl\lbrace 
 R_{10}-\frac 12 \left(\nabla \Phi\right)^2\\
&-\frac 12 e^{-\Phi}\left(\frac{(h_1-h_3)^2}{2\Omega_1^2\Omega_2^2\Omega_3^2}+\frac{1}{\Omega_3^4}\left(\nabla h_1\right)^2
+\frac{1}{\Omega_2^4}\left(\nabla h_3\right)^2\right)
\\
&-\frac 12 e^{\Phi}\left(\frac{2}{\Omega_2^2\Omega_3^2}\left(\nabla h_2\right)^2
+\frac{1}{\Omega_1^2\Omega_2^4}\left(h_2-\frac P9\right)^2
+\frac{1}{\Omega_1^2\Omega_3^4} h_2^2\right)
\\
&-\frac {1}{2\Omega_1^2\Omega_2^4\Omega_3^4}\left(4{\Omega}_0+ h_2\left(h_3-h_1\right)+\frac 19 P h_1\right)^2
\biggr\rbrace\,,\\
\end{split}
\eqlabel{5action}
\end{equation}
where
\begin{equation}
\begin{split}
R_{10}=R_5&+\left(\frac{1}{2\Omega_1^2}+\frac{2}{\Omega_2^2}+\frac{2}{\Omega_3^2}-\frac{\Omega_2^2}{4\Omega_1^2\Omega_3^2}
-\frac{\Omega_3^2}{4\Omega_1^2\Omega_2^2}-\frac{\Omega_1^2}{\Omega_2^2\Omega_3^2}\right)-2\Box \ln\left(\Omega_1\Omega_2^2\Omega_3^2\right)\\
&-\biggl\{\left(\nabla\ln\Omega_1\right)^2+2\left(\nabla\ln\Omega_2\right)^2
+2\left(\nabla\ln\Omega_3\right)^2+\left(\nabla\ln\left(\Omega_1\Omega_2^2\Omega_3^2\right)\right)^2\biggr\}\,,
\end{split}
\eqlabel{ric5}
\end{equation}
is the Ricci scalar of the ten-dimensional background geometry 
\begin{equation}
ds_{10}^2 =g_{\mu\nu}(y) dy^{\mu}dy^{\nu}+\Omega_1^2(y) g_5^2
+\Omega_2^2(y) \left[g_3^2+g_4^2\right]+\Omega_3^2(y) \left[g_1^2+g_2^2\right],
\eqlabel{10met}
\end{equation}
with the  one-forms $g_i$ are the usual forms of  the $T^{1,1}$ (see \cite{Buchel:2010wp} for explicit expressions),
and $R_5$ is the five-dimensional Ricci scalar of
\begin{equation}
ds_{5}^2 =g_{\mu\nu}(y) dy^{\mu}dy^{\nu}\,.
\eqlabel{5met}
\end{equation}
The scalars $h_i=h_i(y)$ parameterize the three forms fluxes, $\Phi=\Phi(y)$ is the dilaton and the
constant\footnote{In the limit of the vanishing 3-form fluxes, $\Omega_0=\frac{L^4}{108}$, where $L$ is the asymptotic
$AdS_5$ radius.}
$\Omega_0$ determine the five-form flux (see \cite{Buchel:2010wp} for the details). 
Parameter $P$ must be appropriately quantized \cite{Herzog:2001xk},
\begin{equation}
P=\frac 92 M\alpha'\,,
\eqlabel{defp}
\end{equation}
corresponding to the integer $M$, \ie the difference of the ranks of the cascading gauge theory gauge group factors. 
 Finally, $G_5$ is the five dimensional effective gravitational constant  
\begin{equation}
G_5\equiv \frac{G_{10}}{{\rm{vol}}_{T^{1,1}}}=\frac{27}{16\pi^3}\ G_{10}\,,
\eqlabel{g5deff}
\end{equation}
where $16 \pi G_{10}=(2\pi)^7(\alpha')^4$ is  10-dimensional gravitational constant of 
type IIB supergravity.

In what follows we introduce 
\begin{equation}
\begin{split}
h_1=&\frac 1P\left(\frac{K_1}{12}-36\Omega_0\right)\,,\qquad h_2=\frac{P}{18}\ K_2\,,\qquad 
h_3=\frac 1P\left(\frac{K_3}{12}-36\Omega_0\right)\,,\\
\Omega_1=&\frac 13 f_c^{1/2} h^{1/4}\,,\qquad \Omega_2=\frac {1}{\sqrt{6}} f_a^{1/2} h^{1/4}\,,\qquad 
\Omega_3=\frac {1}{\sqrt{6}} f_b^{1/2} h^{1/4}\,,\qquad g= e^\Phi\,.
\end{split}
\eqlabel{redef}
\end{equation}

Chirally symmetric 
states of the cascading gauge theory correspond to the enhancement of the
global symmetry  $SU(2)\times SU(2)\times \zet_2\to SU(2)\times SU(2)\times U(1)$,
and are described by the gravitational configurations of \eqref{5action} 
subject to the constraints\footnote{This is a consistent truncation
of the cascading gauge theory to $U(1)$ symmetric
sector constructed in \cite{Aharony:2005zr}.}:   
\begin{equation}
h_1=h_3\,,\qquad h_2=\frac{P}{18}\,,\qquad \Omega_2=\Omega_3\,.
\eqlabel{cinv}
\end{equation}
The chiral symmetry breaking sector is parameterized by 
 $\{\dd f,\dd k_1,\dd k_2\}$ in
\begin{equation}
\begin{split}
K_1=&K+\dd k_1\,,\qquad K_2=1+\dd k_2\,,\qquad K_3=K-\dd k_1\,,
\\
f_c=&f_2\,,\qquad f_a=f_3+\dd f\,,\qquad f_b=f_3-\dd f\,.
\end{split}
\eqlabel{deffl}
\end{equation}
Its linearized fluctuations on top of the $U(1)$-symmetric on-shell backgrounds 
\begin{equation}
\bigg\{ds_5^2\,, K\,, h\,, 
f_2\,, f_3\,, g\bigg\} 
\eqlabel{symmetric}
\end{equation}
are governed by the following effective action\footnote{Since the fluctuations
$\{\delta f,\delta k_1,\delta k_2\}$ break the chiral $U(1)$ symmetry
to $\zet_2$, they can never mix with any $U(1)$-symmetric fluctuations of the
background \eqref{symmetric}. This was also verified explicitly in the analysis
reported in \cite{Buchel:2010wp},
both from the effective action
perspective, and from the perspective of the second-order equations of motion of the
KS black hole.} \cite{Buchel:2010wp}
\begin{equation}
S_{\csb}\bigg[\dd f,\dd k_1,\dd k_2\bigg]
=\frac{1}{16\pi G_5}\int_{\calm_5}\ {\rm vol}_{\calm_5}\ h^{5/4}f_2^{1/2}f_3^2
\biggl\{\call_1+\call_2+\call_3+\call_4+\call_5\biggr\}\,,
\eqlabel{flaction}
\end{equation}
\begin{equation}
\begin{split}
\call_1=&-\frac{(\dd f)^2}{f_3^2}\left(
-\frac{P^2 e^\Phi}{2 f_2 h^{3/2} f_3^2}-\frac{(\nabla K)^2}{8 f_3^2 h P^2 e^\Phi}- \frac{K^2}{2f_2 h^{5/2} f_3^4}
\right)\,,
\end{split}
\eqlabel{call1}
\end{equation}
\begin{equation}
\begin{split}
\call_2=&-\frac{9f_3^2-24 f_2 f_3+4f_2^2}{f_2h^{1/2}f_3^4}\ (\dd f)^2+2\square\frac{(\dd f)^2}{f_3^2}
-\left(\nabla \frac{(\dd f)^2}{f_3^2}\right)^2\\
&-2\nabla\left(\ln h^{1/4}f_3^{1/2}\right)\nabla 
\left(\frac{(\dd f)^2}{f_3^2}\right)+2\nabla\left(\ln f_2^{1/2} h^{5/4}f_3^2\right)\nabla\left(\frac{(\dd f)^2}
{f_3^2}\right)\,,
\end{split}
\eqlabel{call2}
\end{equation}
\begin{equation}
\begin{split}
\call_3=&-\frac {1}{2P^2 e^\Phi}\biggl(\frac {9}{2 f_2 h^{3/2}f_3^2}\ (\dd k_1)^2
+\frac {1}{2h f_3^4} \biggl(2(\nabla K)^2\ (\dd f)^2+f_3^2\ (\nabla\dd k_1)^2
\\
&+4f_3 \dd f\ \nabla K\nabla \dd k_1
\biggr)
\biggr)\,,
\end{split}
\eqlabel{call3}
\end{equation}
\begin{equation}
\begin{split}
\call_4=&\frac{P^2 e^\Phi}{2}\biggl(\frac {2}{9hf_3^2}\ (\nabla \dd k_2)^2
+\frac{2}{f_2h^{3/2}f_3^4}\left(3\ (\dd f)^2+4f_3\ \dd f\dd k_2+f_3^3\ (\dd k_2)^2 \right)
\biggr)\,,
\end{split}
\eqlabel{call4}
\end{equation}
\begin{equation}
\begin{split}
\call_5=&\frac{K}{f_2 h^{5/2} f_3^6}\ \left(f_3^2\ \dd k_1\dd k_2-K\ (\dd f)^2\right)\,.
\end{split}
\eqlabel{call5}
\end{equation}

\subsection{KT black hole and $\csb$ quasinormal modes}\label{lowt}

The Klebanov-Tseytlin black hole \cite{Aharony:2007vg} is a chirally symmetric (see \eqref{cinv})
solution of the effective action \eqref{5action}. The five-dimensional metric is
\begin{equation}
ds_5^2= h^{-1/2} (2x-x^2)^{-1/2}\left(-(1-x)^2\ dt^2+dx_1^2+dx_2^2+dx_3^2\right)+G_{xx} dx^2 \,,
\eqlabel{ktmetric}
\end{equation}
where 
\begin{equation}
\begin{split}
&G_{xx}=\frac{\sqrt{h} f_3^2}{ 2 (x-1) P^2 g^2 (2-x)^2 x^2 (K^2+8 h^2 f_3^2 f_2 (f_2-6 f_3)+2 h f_3^2 P^2 g) }
\biggl(12 P^2 f_3^2 g^2 f_2 h^2\\
&\times(1-x)+f_2 x^2 (2 P^2 f_3^2 g^2
h'^2 -12 P^2 g^2 h^2 f_3'^2+K'^2 h g
+2 P^2 h^2 f_3^2 g'^2) (x-1) (2-x)^2\\&-4 x P^2 f_3 g^2 h f_2 (2-x) (x^2-2 x+2) (h' f_3+4 f_3' h)
+4 x P^2 f_3 (2-x) g^2 h^2 \\
&\times (2 x f_3' (1-x) (2-x)-(x^2-2 x+2) f_3)
f_2'\biggr)\,.
\end{split}
\eqlabel{forgxx}
\end{equation}
The gravitational bulk scalars\footnote{See eq.(2.8)-(2.12) of \cite{Aharony:2007vg} for the corresponding
equations of motion.} $\{K,h,f_2,f_3,g\}$ are functions of the radial coordinate $x\in (0,1)$.
Asymptotically near the boundary ($x\to 0_+$),
\begin{equation}
K=P^2 g_0 \biggl[k_s-\frac 12 \ln x +\sum_{n=1}^\infty\sum_{k} K_{n,k}\ x^{n/2}\ \ln^k x\biggr]\,,
\eqlabel{kkt}
\end{equation}
\begin{equation}
h=\frac{P^2 g_0}{a_0^2} \biggl[\frac 18+\frac{k_s}{4}-\frac 18 \ln x +\sum_{n=1}^\infty\sum_{k} h_{n,k}\ x^{n/2}\ \ln^k x\biggr]\,,
\eqlabel{hkt}
\end{equation}
\begin{equation}
f_2=a_0 \biggl[1 +\sum_{n=1}^\infty\sum_{k} f_{2,n,k}\ x^{n/2}\ \ln^k x\biggr]\,,
\eqlabel{f2kt}
\end{equation}
\begin{equation}
f_3=a_0 \biggl[1 +\sum_{n=1}^\infty\sum_{k} f_{3,n,k}\ x^{n/2}\ \ln^k x\biggr]\,,
\eqlabel{f3kt}
\end{equation}
\begin{equation}
g=g_0 \biggl[1 +\sum_{n=1}^\infty\sum_{k} g_{n,k}\ x^{n/2}\ \ln^k x\biggr]\,,
\eqlabel{gkt}
\end{equation}
characterized by $\{f_{3,2,0},f_{3,3,0},f_{3,4,0},g_{2,0}\}$ (in addition to $P,g_0,a_0$);
while near the horizon ($y\equiv 1-x\to 0_+$),
\begin{equation}
\begin{split}
&K=P^2g_0\ \sum_{n=0}^\infty k_{n}^h\ y^{2n}\,,\qquad h=\frac{P^2g_0}{a_0^2}\ \sum_{n=0}^\infty h_{n}^h\ y^{2n}\,,\qquad f_2=a_0\ \sum_{n=0}^\infty
f_{2,n}^h\ y^{2n}\,,\\
&f_3=a_0\ \sum_{n=0}^\infty
f_{3,n}^h\ y^{2n}\,,\qquad g=g_0\ \sum_{n=0}^\infty g_{n}^h\ y^{2n}\,,
\end{split}
\eqlabel{kthor}
\end{equation}
completely specified with $\{k_0^h,h_0^h,f_{2,0}^h,f_{3,0}^h,f_{3,1}^h,g_0^h\}$. 
The thermodynamics of the KT black hole was discussed in \cite{Aharony:2007vg};
in what follows we will need only the expression for the Hawking temperature
and the strong coupling scale of the theory, see \cite{Aharony:2007vg,Bena:2019sxm}, 
\begin{equation}
T=\sqrt{\frac{a_0}{P^2g_0}}\ \times\ \sqrt{\frac{3f_{3,0}^h-f_{2,0}^h}
{2 \pi^2 f_{3,0}^hh_0^h(f_{3,0}^h+2 f_{3,1}^h)}}\,,\qquad \Lambda=\frac{\sqrt{a_0}}{P^2 g_0}\ e^{-k_s/2} \,.
\eqlabel{temperature}
\end{equation}

Using the KT background \eqref{ktmetric},
and the plane-wave ansatz for the chiral symmetry breaking fluctuations,
\begin{equation}
\delta f=e^{-i\omega t+i k x_3} F\,,\qquad \delta k_1=e^{-i\omega t+i k x_3} \calk_1\,,\qquad
\delta k_2=e^{-i\omega t+i k x_3} \calk_2\,,
\eqlabel{flucw}
\end{equation}
we find from \eqref{flaction} the equations of motion for the QNM wavefunctions
$\{F,\calk_1,\calk_2\}$, all of them being the functions of the radial coordinate $x$,
\begin{equation}
\begin{split}
&0=F''-\left(\frac{2 f_3}{f_3}+\frac{1}{1-x}\right) F'
-\frac{K'}{2h P^2 g f_3}\ \calk_1'-\frac{2 g P^2 G_{xx}}{h^{3/2} f_3 f_2} \calk_2
+\biggl(\frac{(f_3')^2}{f_3^2}-\frac{(K')^2}{2g h f_3^2 P^2}\\
&+\biggl[h^{1/2} \sqrt{2x-x^2}\ \left(\frac{\omega^2}{(1-x)^2}-k^2\right)
+\frac{4 h f_2^2-9 h f_3^2-2 g P^2}{f_3^2 f_2 h^{3/2}}\biggr] G_{xx}
+\frac{2}{(2x-x^2)^2}\biggr) F\,,
\end{split}
\eqlabel{fl1}
\end{equation}
\begin{equation}
\begin{split}
&0=\calk_1''-\left(\frac{2 f_3'}{f_3}+\frac{h'}{h}+\frac{g'}{g}+\frac{1}{1-x}\right) \calk_1'
+\frac{2 K'}{f_3} F'+\frac{2 K g P^2 G_{xx}}{f_3^2 f_2 h^{3/2}} \calk_2
+\biggl(\frac{4 K g P^2 G_{xx}}{h^{3/2} f_2 f_3^3}\\
&-\frac{2 f_3' K'}{f_3^2}\biggr) F
+\biggl[h^{1/2} \sqrt{2x-x^2} \left(\frac{\omega^2}{(1-x)^2}-k^2\right)-\frac{9}{h^{1/2} f_2}
\biggr] G_{xx} \calk_1\,,
\end{split}
\eqlabel{fl2}
\end{equation}
\begin{equation}
\begin{split}
&0=\calk_2''-\left(\frac{2 f_3'}{f_3}+\frac{h'}{h}-\frac{g'}{g}+\frac{1}{1-x}\right)
\calk_2'+\biggl[h^{1/2} \sqrt{2x-x^2}\left(\frac{\omega^2}{(1-x)^2}-k^2\right)
-\frac{9}{h^{1/2} f_2}\biggr]\\
&\times G_{xx} \calk_2
-\frac{18 G_{xx}}{h^{1/2} f_3 f_2} F+\frac{9K G_{xx}}{2h^{3/2} g f_2 f_3^2 P^2} \calk_1\,.
\end{split}
\eqlabel{fl3}
\end{equation}
To determine the spectrum of the QNMs, one needs to solve \eqref{fl1}-\eqref{fl3} with
``normalizable only'' conditions at the bulk gravitational boundary, and the incoming-wave
boundary conditions are the horizon \cite{Kovtun:2005ev}. Fixing (without the loss of generality)
the normalizable coefficient of $F$ to one, see \eqref{uvfls} below,
the above boundary conditions determine the spectrum of the QNMs:
\begin{equation}
\omega=\omega\left(q\equiv k^2,\frac{T}{\Lambda}\right)\,.
\eqlabel{wq}
\end{equation}
As in \cite{Buchel:2010wp}, we are not interested in the numerical data for the spectrum per se,
rather, we would like to identify QNMs at the threshold of instability.
To this end, we solve
\begin{equation}
0=\omega\left(q,\frac{T}{\Lambda}\right)\bigg|_{q=\frac{a_0}{P^2 g_0}\ \qnm}\qquad \Longrightarrow \qquad
\qnm=\qnm\left(\frac{T}{\Lambda}\right)\,,
\eqlabel{thofin}
\end{equation}
where we introduced the dimensionless $\qnm$.
The KT black hole is unstable with respect to the chiral symmetry breaking fluctuations
for all temperatures $T$ such that  $\qnm>0$  \cite{Buchel:2010wp}, \ie 
\begin{equation}
\qnm>0\qquad \Longleftrightarrow\qquad {\rm KT\ black\ hole}\ \csb\ {\rm perturbative\ instability} 
\eqlabel{inscri}
\end{equation}
Focusing on the instability criteria \eqref{thofin}, the asymptotic expansions for the
linearized $\csb$ fluctuations take form:
\nxt in the UV, \ie as $x\to 0_+$,
\begin{equation}
\begin{split}
&F=a_0\ x^{3/4} \biggl(
f_{3,0}+x^{1/2} \left(-\frac{\sqrt{2}}{32}  \qnm f_{3,0} \ln x+\frac{\sqrt{2}}{96}  \qnm \left(3 f_{3,0} k_s+13 f_{3,0}
-2 k_{2,3,0}\right)\right)\\
&+\calo(x\ln^2 x)\biggr)\,,
\end{split}
\eqlabel{uvfl1}
\end{equation}
\begin{equation}
\begin{split}
&\calk_1=P^2 g_0\ x^{3/4} \biggl(
\frac12 f_{3,0} \ln x+\frac23 f_{3,0}+\frac23 k_{2,3,0}+x^{1/2} \biggl(-\frac{\sqrt{2}}{128}  \qnm f_{3,0} \ln^2 x
+\frac{\sqrt{2}}{192} \qnm (3 f_{3,0} k_s\\
&+f_{3,0}-2 k_{2,3,0}) \ln x
-\frac{\sqrt{2}}{384} \qnm (f_{3,0} k_s-8 k_{2,3,0} k_s-56 f_{3,0}-2 k_{2,3,0})\biggr)
+\calo(x\ln^3 x)\biggr)\,,
\end{split}
\eqlabel{uvfl2}
\end{equation}
\begin{equation}
\begin{split}
&\calk_2=x^{3/4} \biggl(
\frac34 f_{3,0} \ln x+k_{2,3,0}+x^{1/2} \biggl(-\frac{3\sqrt{2}}{256}  \qnm f_{3,0} \ln^2 x
+\frac{\sqrt{2}}{128} \qnm (3 f_{3,0} k_s+3 f_{3,0}-2 k_{2,3,0})\\
&\times\ln x
+\frac{\sqrt{2}}{640}  \left(-\frac{15}{2} \qnm k_s f_{3,0}+20 k_{2,3,0} k_s \qnm
+75 \qnm f_{3,0}+15 k_{2,3,0} \qnm\right)\biggr)\\
&+x \biggl(k_{2,7,0}+\cdots+\calo(\ln^3x )\biggr)
+\calo(x^{3/2}\ln^4 x)\biggr)\,,
\end{split}
\eqlabel{uvfl3}
\end{equation}
characterized by (we fixed the overall normalization of the linearized fluctuations
setting the normalizable coefficient of $F$)
\begin{equation}
\{f_{3,0}=1\,,\ k_{2,3,0}\,,\ k_{2,7,0}\,,\ \qnm\}\,;
\eqlabel{uvfls}
\end{equation}
\nxt in the IR, \ie as $y=1-x\to 0_+$,
\begin{equation}
F=a_0\biggl(F_0^h+\calo(y^2)\biggr)\,,\qquad
\calk_1=P^2g_0\biggl(K_{1,0}^h+\calo(y^2)\biggr)\,,\qquad \calk_2=K_{2,0}^h+\calo(y^2)\,,
\eqlabel{irfl}
\end{equation}
characterized (in addition to $\qnm$) by 
\begin{equation}
\{F_{0}^h\,,\ K_{1,0}^h\,,\ K_{2,0}^h\}\,.
\eqlabel{irfls}
\end{equation}

Given a KT black hole solution at a certain temperature, we expect
three distinct branches of the quasinormal modes --- the $\csb$ sector mixes in three gravitational modes
$\{\delta f, \delta k_1, \delta k_2\}$, dual to a pair of dimension $\Delta=3$ operators
(the normalizable coefficients $f_{3,0}$ and $k_{2,3,0}$), and a single dimension $\Delta=7$ operator
(the normalizable coefficient $k_{2,7,0}$). Furthermore, each QNM branch has its own tower of
excitations, with increasingly higher $\qnm$, characterized by the number of nodes in the
radial wavefunction. In section \ref{hight} we identify each QNM  branch at high temperatures,
where particular linear combinations of $\{\delta f, \delta k_1, \delta k_2\}$ decouple.
We then follow the lowest QNM of each branch numerically to low temperatures.
We employ two different computational schemes: SchemeA and SchemeB.
\begin{itemize}
\item SchemeA: as in \cite{Buchel:2010wp}, we set $P=a_0=g_0=1$ as vary $k_s$ --- this is a convenient regime
to reach low temperatures.
\item SchemeB: alternatively, we set $a_0=g_0=1$ and $k_s=\frac 1b$, $P^2=b$ --- this is a convenient regime
to reach high temperatures.
\end{itemize}
Results of these
computations are collected in  Figs.~\ref{figure33ahigh}-\ref{figure3a7}.
We plot dimensionless quantities:
\begin{equation}
\kk^2\equiv \frac{k^2}{(2\pi T)^2}=\frac{q}{(2\pi T)^2}=\frac{a_0}{P^2g_0}\times \frac{\qnm}{(2\pi T)^2}=
\qnm\ \frac{f_{3,0}^hh_0^h(f_{3,0}^h+2 f_{3,1}^h)}{2(3f_{3,0}^h-f_{2,0}^h)}\,,
\eqlabel{defqq}
\end{equation}
as a function of
\begin{equation}
\frac{T}{\Lambda}=\sqrt{P^2 g_0}\ e^{ks/2}\ \sqrt{\frac{3f_{3,0}^h-f_{2,0}^h}
{2 \pi^2 f_{3,0}^hh_0^h(f_{3,0}^h+2 f_{3,1}^h)}}\,,
\eqlabel{deftl}
\end{equation}
in the computational SchemeA, 
or as a function of $b$ in the computational SchemeB.

\subsection{High-temperature (near-conformal) limit}\label{hight}

There are three branches of the quasinormal modes associated with the $\csb$ sector of the cascading gauge theory.  
In this section we explain how these branches are identified by the decoupled linear
combinations of the $\csb$ fluctuations
$\{\delta f,\delta k_1,\delta k_2 \}$ in the KT black hole background at high temperatures. 
We use the computational SchemeB (see section \ref{hight}), \ie the KT black hole in the near conformal
$b\to 0$ limit.

Exactly at $b=0$ the KT BH is just the $AdS_5$-Schwarzschild black hole:
\begin{equation}
h\equiv\frac 14\,,\qquad K\equiv 1\,,\qquad  f_{2}=f_3\equiv 1\,,\qquad g\equiv 1\,,
\eqlabel{b0limit}
\end{equation}
where the $AdS_5$ radius is 
\begin{equation}
L^4=\frac 14\,.
\eqlabel{lpert}
\end{equation}
The leading order $\calo(b)$ corrections were discussed in \cite{Gubser:2001ri}; here we follow the state-of-the-art
construction of \cite{Buchel:2009bh}, done to $\calo(b^4)$ inclusive:
\begin{equation}
\begin{split}
&h(x)=\frac 14+\sum_{n=1}^\infty\bigg\{\ b^n\ \left(\xi_{2n}(x)-\frac 54\eta_{2n}(x)\right)\ \bigg\}\,,\\
&f_2(x)=1+\sum_{n=1}^\infty\bigg\{\ b^n\ \left(-2\xi_{2n}(x)+\eta_{2n}(x)+\frac 45\lambda_{2n}(x)\right)\ \bigg\}\,,\\
&f_3(x)=1+\sum_{n=1}^\infty\bigg\{\ b^n\ \left(-2\xi_{2n}(x)+\eta_{2n}(x)-\frac 15\lambda_{2n}(x)\right)\ \bigg\}\,,\\
&K(x)=1+\sum_{n=1}^\infty\bigg\{\ b^n\ \kappa_{2n}(x)\ \bigg\}\,,\\
&g(x)=1+\sum_{n=1}^\infty\bigg\{\ b^n\ \zeta_{2n}(x)\ \bigg\}\,.
\end{split}
\eqlabel{pertkt}
\end{equation}
The equations for $\{\kappa_{2n},\xi_{2n},\eta_{2n},\lambda_{2n},\zeta_{2n}\}$ decouple at each order
$n$, see eqs.~(2.16)-(2.20) of  \cite{Buchel:2009bh}. At order $n=1$, $\kappa_2$ and $\xi_2$
can be determined analytically\footnote{Any free integration
constants are fixed so to enforce the computational SchemeB.}:
\begin{equation}
\kappa_2=-\frac 12 \ln x -\frac 12\ln\left(1-\frac 12 x\right)\,,\qquad
\xi_2=\frac{1}{12} (\ln x-1)+\frac{1}{12}\ln\left(1-\frac 12 x\right)\,,
\eqlabel{k2xi2}
\end{equation}
while the remaining functions $\eta_2,\lambda_2,\zeta_2$  has to be determined numerically.
We will need the asymptotes of these functions:
\nxt in the UV, \ie as $x\to 0_+$,
\begin{equation}
\begin{split}
&\eta_2=\frac 16\ln x-\frac 16-\frac{x}{30}+x^2\left(\eta_{2,2,0}+\frac{1}{30}\ln x\right)
+\calo(x^3\ln x)\,,\\
&\lambda_2=\frac 23 x +\lambda_{2,3} x^{3/2}+\calo(x^2)\,,\qquad
\zeta_2=x\left(\frac 12\ln x +\zeta_{2,1,0}\right)+\calo(x^2\ln x)\,;
\end{split}
\eqlabel{uvpert}
\end{equation}
\nxt in the IR, \ie as $y=1-x\to 0_+$,
\begin{equation}
\begin{split}
&\eta_2=\eta_{2,0}^h+\left(\frac{7}{60}+\frac 13\ln 2+2\eta_{2,0}^h\right)y^2+\calo(y^4)\,,\\
&\lambda_2=\lambda_{2,0}^h+\left(\frac 34\lambda_{2,0}^h-\frac 14\right)y^2+\calo(y^4)\,,\qquad
\zeta_2=\zeta_{2,0}^h+\frac 14y^2 +\calo(y^4)\,.
\end{split}
\eqlabel{irpert}
\end{equation}
To order $\calo(b)$ the Hawking temperature of the KT black hole can be computed analytically,
\begin{equation}
2\pi T=4-(1+\ln 2)b+\calo(b^2)\,,
\eqlabel{tpert}
\end{equation}
additionally, from \eqref{deftl},
\begin{equation}
\frac 1b+\calo(\ln b) = 2\ln \frac{T}{\Lambda}\,.
\eqlabel{smallb}
\end{equation}

Assuming, at fixed $\omega$, 
\begin{equation}
\begin{split}
&F(x)=\sum_{n=1}^\infty \bigg\{\ b^{n/2}\ F_n(x)\ \bigg\}\,,\\
&\calk_1(x)=b\ \sum_{n=0}^\infty \bigg\{\ b^{n/2}\ \left(\frac 13\calk_{3,n}(x)-\frac 13 \calk_{7,n}(x)\right)
\ \bigg\}\,,\\
&\calk_2(x)=\sum_{n=0}^\infty \bigg\{\ b^{n/2}\ \left(\frac 12\calk_{3,n}(x)+\frac 12 \calk_{7,n}(x)\right)
\ \bigg\}\,,\\
&k^2\equiv q(\omega)=\sum_{n=0}^\infty\  \bigg\{\ b^{n/2}\ q_n(\omega)\ \bigg\}\,,
\end{split}
\eqlabel{flpert}
\end{equation}
we find from \eqref{fl1}-\eqref{fl3} that at each order $n\ge 1$ the equations of motion for $\{F_n,\calk_{3,n-1},\calk_{7,n-1}\}$ decouple:
\begin{equation}
\begin{split}
0=&F_n''-\frac{F_n'}{1-x} +\frac{1}{16 (2x-x^2)^{3/2}}
\left(\frac{\omega^2}{(1-x)^2}-q_0\right) F_n+\frac{3F_n}{4(2x-x^2)^2}+\calj_{F}^{[n]}\,,
\end{split}
\eqlabel{fpert}
\end{equation}
\begin{equation}
\begin{split}
0=&\calk_{3,n}''-\frac{\calk_{3,n}'}{1-x}+\frac{1}{16 (2x-x^2)^{3/2}}
\left(\frac{\omega^2}{(1-x)^2}-q_0\right) \calk_{3,n}+\frac{3 \calk_{3,n}}{4(2x-x^2)^2}\,,
+\calj_{3}^{[n]}
\end{split}
\eqlabel{3pert}
\end{equation}
\begin{equation}
\begin{split}
0=&\calk_{7,n}''-\frac{\calk_{7,n}'}{1-x}+
\frac{1}{16(2x-x^2)^{3/2}}
\left(\frac{\omega^2}{(1-x)^2}-q_0\right) \calk_{7,n}-\frac{21 \calk_{7,n}}{4(2x-x^2)^2}+
\calj_7^{[n]}\,,
\end{split}
\eqlabel{7pert}
\end{equation}
where the source terms $\{\calj_{F}^{[n]},\calj_{3}^{[n]},\calj_{7}^{[n]}\}$ are functionals
of the lower order solutions, $\{F_m,\calk_{3,m},\calk_{7,m}\,;\, \xi_{2m},\eta_{2m},
\kappa_{2m},\zeta_{2m}\}$ and $q_m$, with $m< n$; additionally
$\{\calj_{3}^{[n]},\calj_{7}^{[n]}\}$ also depend on $F_n$.

The source functionals $\calj_F^{[1]}$, $\calj_{3}^{[0]}$ and $\calj_{7}^{[0]}$
identically vanish. Consider a 
free, minimally coupled gravitational bulk scalar $\phi$  in $AdS_5$,
dual to an operator of  dimension  $\Delta$  of the boundary conformal theory.
In $AdS_5$-Schwarzschild black hole background \eqref{b0limit} the corresponding QNM equation
takes form\footnote{Solving this equation with $k=0$ we reproduce the QNM spectra of
$AdS_5$-Schwarzschild reported in \cite{Nunez:2003eq}.}
\begin{equation}
0=\phi''-\frac{\phi'}{1-x} +\frac{1}{16(2x-x^2)^{3/2}}
\left(\frac{\omega^2}{(1-x)^2}-k^2\right) \phi-\frac{\Delta(\Delta-4)\phi}{4(2x-x^2)^2}\,.
\eqlabel{qnmphi}
\end{equation}
Thus, we identify $F_1$ and $\calk_{3,0}$ with the gravitational duals to a
pair of the dimension $\Delta=3$ operators,
and $\calk_{7,0}$ with the gravitational dual to the dimension $\Delta=7$ operator. 

We can now identify distinct branches of the quasinormal modes of the KT black hole:
\nxt The $\calb_7$ branch is defined by the boundary conditions
\begin{equation}
\calk_{3,0}(x)\equiv 0\,,\qquad
\calk_{7,0}(x)=x^{7/4}\left(1+\calo(x^{1/2})\right)\,,
\eqlabel{defb7}
\end{equation}
where we fixed to unity the normalizable coefficient of $\calk_{7,0}$.
This branch is analytic in the conformal deformation parameter $b$:
for all $k\ge 0$, $F_{2k+1}$, $\calk_{3,2k+1}$ and $\calk_{7,2k+1}$ vanish identically.   
We construct the $\calb_7$ branch to order $\calo(b)$ in appendix \ref{app7}.
\nxt A pair of $\Delta=3$ QNM branches is defined by the boundary conditions
\begin{equation}
\calk_{3,0}(x)=x^{3/4}\left(1+\calo(x^{1/2})\right)\,,\qquad
\calk_{7,0}(x)\equiv 0\,,\qquad F_1=\alpha_0\ \calk_{3,0}\,,
\eqlabel{defb3}
\end{equation}
where $\alpha_0\ne 0$ is a constant, which is fixed at the subleading order. 
As in \eqref{defb7}, we fixed to unity the normalizable coefficient of $\calk_{3,0}$.
As we explain in appendix \ref{app3}, there are two possible choices of $\alpha_0$, differing by the overall sign --- this leads to two possible values of $q_1$,
\begin{equation}
q=q_0\pm |q_1|\ b^{1/2}+\calo(b)\,.
\eqlabel{qdim3}
\end{equation}
We call the '$+$' branch in \eqref{qdim3} $\calb_{3u}$ ---
as we follow this branch to low temperatures,
it identifies QNMs spontaneously breaking the chiral symmetry\footnote{This is the
branch of the QNMs originally found in \cite{Buchel:2010wp}.}. The other branch,
the '-' branch in \eqref{qdim3} --- we call it $\calb_{3s}$ --- represents the stable
quasinormal modes.

\section{KS branches from explicit $\csb$ fluctuations of the KT black hole}\label{sussec}

There is an alternative approach to identify spontaneous symmetry broken phases in holographic duals
advocated in \cite{Buchel:2019pjb}. Unlike the analysis of the QNMs in the symmetry breaking sector, it does no
identify exactly what mode becomes unstable, but it does determine the onset of the instability. 
The idea is simple. The homogeneous and isotropic ``branches'' of the symmetry-broken KS black hole connect to the
KT black hole ``trunk" wherever the parameters of the latter allow for a linearized normalizable fluctuations in the
symmetry breaking sector. If, exactly at the onset of the instability, one turns on a non-normalizable coefficient
for the fluctuations, their normalizable coefficients will necessarily diverge. Thus, a way to identify
onset of the $\csb$ instabilities of the KT black hole is to monitor for the divergence
of the expectation values of the condensates (such as parameters $\{f_{3,0},k_{2,3,0},k_{2,7,0}\}$ in  \eqref{uvfl1}-\eqref{uvfl3} ) as the KT hole temperature varies, provided we turn on the non-normalizable coefficient ---
here, a gravitational dual to one of the  gaugino mass parameters \cite{Buchel:2010wp}. 

The relevant set of equations is \eqref{fl1}-\eqref{fl3} with $\omega=k=0$.
We use the computational SchemeA  (see section \ref{hight}),
\ie we set  $P=a_0=g_0=1$ and vary $k_s$ parameter of the KT black hole. 
It is possible to turn on two independent  non-normalizable coefficients $\mu_1$ and $\mu_2$
in the chiral symmetry breaking sector:
\begin{equation}
\begin{split}
&F=\left(\mu_1 (k_s+2)-\left(\frac12 \mu_1+\mu_2\right) \ln x\right) x^{1/4}
+f_{3,0} x^{3/4}+\biggl(
\frac{249}{2} f_{3,2,0} \mu_1+83 \mu_2 f_{3,2,0}\\
&-\frac14 g_{2,0} \mu_1-\frac12 \mu_2 g_{2,0}+\frac18 \mu_1 k_s
+\frac{77}{2} f_{3,2,0} \mu_1 k_s-6 \mu_2 f_{3,2,0} k_s-\frac12 \mu_2+\biggl(
-\frac{71}{4} f_{3,2,0} \mu_1\\
&-\frac{71}{2} \mu_2 f_{3,2,0}-\frac{1}{16} \mu_1-\frac18 \mu_2\biggr) \ln x
\biggr) x^{5/4}+\biggl(f_{7,0}+\biggl(-6 f_{3,0} f_{3,2,0}-\frac32 f_{3,3,0} \mu_1\\&-3 f_{3,3,0}
\mu_2\biggr)\ln x\biggr) x^{7/4}+\calo(x^{9/4}\ln^3 x)\,,
\end{split}
\eqlabel{flm1}
\end{equation}
\begin{equation}
\begin{split}
&\calk_1=\mu_2 (k_s+2) x^{1/4}+\left(k_{1,3,0}+\frac12  f_{3,0} \ln x\right) x^{3/4}+\calo(x^{5/4}\ln^2 x)\,,
\end{split}
\eqlabel{flm2}
\end{equation}
\begin{equation}
\begin{split}
&\calk_2=\left(-\frac32 \mu_1 k_s+\frac32 k_s \mu_2-\frac{15}{4} \mu_1+\frac32 \mu_2
+\left(\frac34 \mu_1+\frac32 \mu_2\right) \ln x\right) x^{1/4}+\biggl(
-f_{3,0}+\frac32 k_{1,3,0}\\
&+\frac34 f_{3,0}\ln x\biggr) x^{3/4}+\calo(x^{5/4}\ln^2 x)\,,
\end{split}
\eqlabel{flm3}
\end{equation}
where the normalizable near the boundary, \ie as $x\to 0_+$, coefficients are
\begin{equation}
\{f_{3,0}\,,\, k_{1,3,0}\,,\, f_{7,0}\}\,.
\eqlabel{massn}
\end{equation}
In the IR we have the asymptotic expansion as in \eqref{irfls}.

\begin{figure}[t]
\begin{center}
\psfrag{f}[cc][][0.8][0]{$f_{3,0}^{-1}\,,\, {\color{red} k_{1,3,0}}^{-1}\,,\, {\color{green} f_{7,0}}^{-1}$}
\psfrag{t}[cc][][1][0]{$\frac{T}{\Lambda}$}
\includegraphics[width=3in]{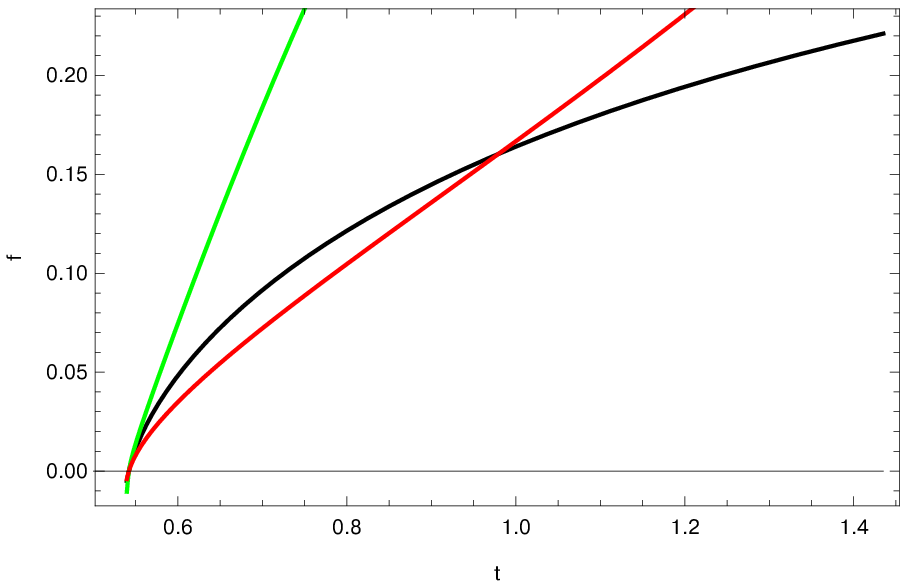}\,
\includegraphics[width=3in]{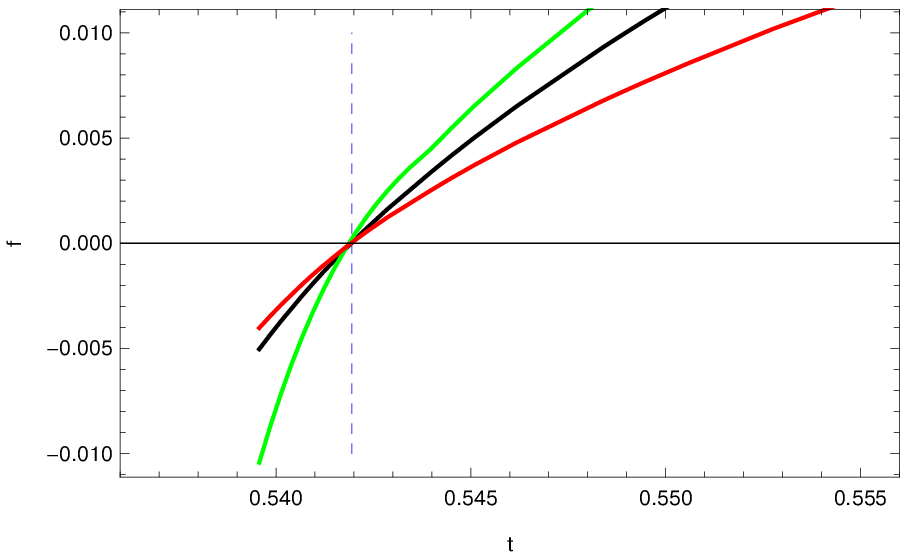}
\end{center}
  \caption{Inverses of the normalizable coefficients
  $f_{3,0}\,,\, {\color{red} k_{1,3,0}}\,,\, {\color{green} f_{7,0}}$ (see \eqref{flm1}-\eqref{flm3})
  of the static homogeneous linearized chiral symmetry breaking
  fluctuations about the KT black hole background as functions of the KT black hole temperature.
  The nonnormalizable coefficients are set as  $\mu_1=0$ and $\mu_2=1$. 
All the normalizable coefficients diverge at $T=T_\csb$ (a vertical dashed blue line in the right panel);
there are no additional divergences for $T>T_{\csb}$.
} \label{figuresus}
\end{figure}

Precisely how we turn on the non-normalizable parameters is irrelevant --- what matters is that $\mu_i$'s are not
simultaneously zero. We set $\mu_1=0$ and $\mu_2=1$. The numerical results for the normalizable
coefficients \eqref{massn} are presented in fig.~\ref{figuresus}. All the normalizable coefficients diverge
at $T=T_\csb\equiv 0.54195 \Lambda$, denoted by the vertical dashed blue line (see the right panel) --- this is the 
$\csb$ instability identified in \cite{Buchel:2010wp}. There are no other divergences for $T>T_\csb$. 
Once again, we conclude that there are no additional KS black hole branches beyond the one identified in \cite{Buchel:2010wp} and constructed in  \cite{Buchel:2018bzp}.

\section*{Acknowledgments}
Research at Perimeter
Institute is supported by the Government of Canada through Industry
Canada and by the Province of Ontario through the Ministry of
Research \& Innovation. This work was further supported by
NSERC through the Discovery Grants program.

\appendix

\section{$\calb_{7}$  branch of the QNMs of the KT black hole
in the $b\to 0$ conformal limit}\label{app7}

We construct here $\calb_{7}$ branch to order $\calo(b)$.
The relevant equations of motion at $\omega=0$ are
\begin{equation}
\begin{split}
&0=\calk_{7,0}''-\frac{\calk_{7,0}'}{1-x}-
\frac{q_0\calk_{7,0}}{16 (2x-x^2)^{3/2}} -\frac{21 \calk_{7,0}}{4(2x-x^2)^2}\,,
\end{split}
\eqlabel{br71}
\end{equation}
at order $\calo(b^0)$, and 
\begin{equation}
\begin{split}
&0=F_{2}''-\frac{F_2'}{1-x}-
\frac{q_0 F_2}{16 (2x-x^2)^{3/2}}+\frac{3 F_2}{4(2x-x^2)^2 x^2}+
\frac 23 \kappa_2' \calk_{7,0}'-\frac{\calk_{7,0}}{(2x-x^2)^2}\,,
\end{split}
\eqlabel{br72}
\end{equation}
\begin{equation}
\begin{split}
&0=\calk_{7,2}''-\frac{\calk_{7,2}'}{1-x}-
\frac{q_0\calk_{7,2}}{16 (2x-x^2)^{3/2}} -\frac{21 \calk_{7,2}}{4(2x-x^2)^2}
-\frac{\calk_{7,0} q_2}{16(2x-x^2)^{3/2}}-3 \kappa_2' F_2'\\
&-\frac{21F_2}{2(2x-x^2)^2}+\left(3 \eta_2'+\frac25 \lambda_2'\right) \calk_{7,0}'+\biggl(
\left(\frac{q_0}{48} \sqrt{2x-x^2}+\frac74\right) (\kappa_2')^2-\frac{(x^2-2 x+2)}{(2x-x^2) (1-x)}\\
&\times\left(\frac {q_0}{8}\sqrt{2x-x^2}+\frac{21}{2}\right) \xi_2'-\frac{1}{48 (2x-x^2)^2} (
480 \kappa_2+1440 \eta_2-144 \lambda_2+84+q_0\sqrt{2x-x^2}\\
&\times(4 \kappa_2+6 \xi_2+1)
\biggr) \calk_{7,0}\,,
\end{split}
\eqlabel{br73}
\end{equation}
\begin{equation}
\begin{split}
&0=\calk_{3,2}''-\frac{\calk_{3,2}'}{1-x}-
\frac{q_0\calk_{3,2}}{16 (2x-x^2)^{3/2}} +\frac{3 \calk_{3,2}}{4(2x-x^2)^2}+\zeta_2' \calk_{7,0}'+3 \kappa_2' F_2'
+\frac{3F_2}{2(2x-x^2)^2}\\&+\frac{3 \zeta_2 \calk_{7,0}}{(2x-x^2)^2}\,,
\end{split}
\eqlabel{br74}
\end{equation}
at order $\calo(b^1)$.
Additionally, $F_1(x)=\calk_{3,0}(x)=\calk_{3,1}(x)=\calk_{7,1}(x)\equiv 0$ and $q_1=0$.

The asymptotic expansions near the boundary, \ie as $x\to 0_+$,
\begin{equation}
\begin{split}
&\calk_{7,0}= x^{7/4}\ \left(1+\frac{\sqrt{2}}{96} q_0\  x^{1/2}+\calo(x)\right)\,,
\end{split}
\eqlabel{br7uv1}
\end{equation}
\begin{equation}
\begin{split}
&F_{2}=x^{3/4} \left(f_{2;3,0}+\frac{\sqrt{2}}{32}  q_0  f_{2;3,0}\ x^{1/2}+\calo(x)\right)\,,
\end{split}
\eqlabel{br7uv2}
\end{equation}
\begin{equation}
\begin{split}
&\calk_{7,2}=x^{3/4} \biggl(-f_{2;3,0}-\frac{ \sqrt{2}}{128} q_0 f_{2;3,0}\  x^{1/2}
+ \biggl(
k_{7,2;7,0}-\left(\frac12-\frac{9f_{2;3,0}}{20} +\frac{q_0^2 f_{2;3,0}}{10240} \right) \ln x
\biggr)\ x\\
&+x^{3/2} \biggl(
\frac{\sqrt{2}}{96} q_2-\left(
\frac{ q_0^3 \sqrt{2}}{983040} f_{2;3,0}-\frac{3 \sqrt{2}}{640} q_0  f_{2;3,0}+\frac{\sqrt{2}}{96} q_0 \right) \ln x
+\frac{\sqrt{2}}{1474560} q_0^3  f_{2;3,0}\\
&+\frac{517 \sqrt{2}}{15360} q_0 f_{2;3,0}+\frac{\sqrt{2}}{96} k_{7,2;7,0} q_0+\frac{55 \sqrt{2}}{3456}
q_0 \biggr)
+\calo(x^2\ln x) \biggr)\,,
\end{split}
\eqlabel{br7uv3}
\end{equation}
\begin{equation}
\begin{split}
&\calk_{3,2}=x^{3/4} \biggl(
\frac 32 f_{2;3,0} \ln x+k_{3,2;3,0}+x^{1/2} \left(\frac{3 \sqrt{2}}{64}  f_{2;3,0} \ln x-\frac{3 \sqrt{2}}{64}   f_{2;3,0}
+\frac{\sqrt{2}}{32}
 k_{3,2;3,0}
\right) q_0
\\&+\calo(x\ln x)
\biggr)\,,
\end{split}
\eqlabel{br7uv4}
\end{equation}
are characterized by
\begin{equation}
\left\{q_0\,,\, f_{2;3,0}\,,\, q_2\,,\, k_{7,2;7,0}\,,\, k_{3,2;3,0}\right\}\,.
\eqlabel{b7para}
\end{equation}
As part of the overall normalization, we can set
\begin{equation}
k_{7,2;7,0}=0\,.
\eqlabel{fixk7270}
\end{equation}
Near the black hole horizon, \ie as $y\equiv 1-x\to 0_+$,
\begin{equation}
\begin{split}
&\calk_{7,0}=k_{7,0;0}^h+\calo(y^2)\,,\qquad F_{2}=f_{2;0}^h+\calo(y^2)\,,\\
&\calk_{7,2}=k_{7,2;0}^h+\calo(y^2)\,,\qquad
\calk_{3,2}=k_{3,2;0}^h+\calo(y^2)\,,
\end{split}
\eqlabel{br7ir}
\end{equation}
characterized by
\begin{equation}
\left\{k_{7,0;0}^h\,,\, f_{2;0}^h\,,\, k_{7,2;0}^h\,,\, k_{3,2;0}^h\right\}\,.
\eqlabel{b7parair}
\end{equation}
In total we have $5+4-1=8$ (from \eqref{b7para} and \eqref{br7ir} minus the constraint \eqref{fixk7270})
parameters
to numerically solve the system of four second-order ODEs \eqref{br71}-\eqref{br74}. 
Discrete solutions are characterized by the number of nodes in the radial wavefunction $\calk_{7,0}$ ---
this is the tower of the excited QNMs on the $\calb_7$ branch. We focus on the lowest QNM on the branch, \ie
we require that $\calk_{7,0}$ does not have nodes. For the lowest QNM on the $\calb_7$ branch we find:
\begin{center}
\[
\begin{tabular}{|c|c|c|c|c|c|c|c|}
\hline
$q_0$  &  $f_{2;3,0}$ & $q_2$  & $k_{3,2;3,0}$ & $k_{7,0;0}^h$ &  $f_{2;0}^h$ &  $k_{7,2;0}^h$ & $k_{3,2;0}^h$\\
\hline\hline
$-164.395$  &  $0.153$ & $185.391$  & $0.383$ & $0.0828$ &  $0.0297$ &  $-0.200$ & $-0.0720$\\
\hline
\end{tabular}
\]
\[
\]

\end{center}

Note that
\begin{equation}
\kk^2\equiv \frac{q}{(2\pi T)^2}=\frac{q_0}{16}+\frac{1}{32}\biggl((1+\ln 2)\ q_0+2\ q_2\biggr)\ b+\calo(b^2) \,,
\eqlabel{br7q2}
\end{equation}
where we used \eqref{tpert}.
The high-temperature approximation \eqref{br7q2} to the
QNM branch $\calb_7$ is shown as a dashed green line on the right panel
of fig.~\ref{figure7high}.

\section{$\calb_{3u}$ and $\calb_{3s}$ branches of the KT black hole QNMs in the $\sqrt{b}\to 0$ conformal limit}\label{app3}

We construct here $\calb_{3u}$ and $\calb_{3s}$ branches to order $\calo(\sqrt{b})$.
The relevant equations of motion at $\omega=0$ are
\begin{equation}
\begin{split}
&0=\calk_{3,0}''-\frac{\calk_{3,0}'}{1-x}-
\frac{q_0\calk_{3,0}}{16 (2x-x^2)^{3/2}} =\frac{3 \calk_{3,0}}{4(2x-x^2)^2}\,,
\end{split}
\eqlabel{br31}
\end{equation}
at order $\calo(b^0)$, and 
\begin{equation}
\begin{split}
&0=F_{2}''-\frac{F_2'}{1-x}-
\frac{q_0 F_2}{16 (2x-x^2)^{3/2}}+\frac{3 F_2}{4(2x-x^2)^2 x^2}-
\frac 23 \kappa_2' \calk_{3,0}'-\frac{\calk_{3,0}}{(2x-x^2)^2}\\
&-\frac{\alpha_0 q_1\calk_{3,0}}{16(2x-x^2)^{3/2}}\,,
\end{split}
\eqlabel{br32}
\end{equation}
\begin{equation}
\begin{split}
&0=\calk_{3,1}''-\frac{\calk_{3,1}'}{1-x}-
\frac{q_0\calk_{3,1}}{16 (2x-x^2)^{3/2}} +\frac{3 \calk_{3,1}}{4(2x-x^2)^2}+3\alpha_0 \kappa_2' \calk_{3,0}'
+\frac{3\alpha_0\calk_{3,0}}{2(2x-x^2)^2}\\&-\frac{q_1\calk_{3,0}}{16(2x-x^2)^{3/2}}\,,
\end{split}
\eqlabel{br34}
\end{equation}
\begin{equation}
\begin{split}
&0=\calk_{7,1}''-\frac{\calk_{7,1}'}{1-x}-
\frac{q_0\calk_{7,1}}{16 (2x-x^2)^{3/2}} -\frac{21 \calk_{7,1}}{4(2x-x^2)^2}
-3\alpha_0 \kappa_2' \calk_{3,0}'-\frac{21\alpha_0\calk_{3,0}}{2(2x-x^2)^2}\,,
\end{split}
\eqlabel{br33}
\end{equation}
at order $\calo(\sqrt{b})$.
Additionally, $F_1(x)\equiv\alpha_0 \calk_{3,0}(x)\,,\, \calk_{7,0}(x)\equiv 0$.
Note that \eqref{br32}-\eqref{br33} have a $\zet_2$ symmetry:
\begin{equation}
\{F_2(x)\,,\,\calk_{3,1}(x)\,,\,\calk_{7,1}(x)\,,\,\alpha_0,q_1\}\ \Longleftrightarrow\
\{F_2(x)\,,\,-\calk_{3,1}(x)\,,\,-\calk_{7,1}(x)\,,\,-\alpha_0,-q_1\}\,.
\eqlabel{zet2sym}
\end{equation}
Both equations \eqref{br32} and \eqref{br34} have zero modes: if $\{F_2,\calk_{3,1}\}$ are solutions, for arbitrary constants $\alpha_1$ and $\beta$,
\begin{equation}
F_2\ \to\ F_2+\alpha_1\ \calk_{3,0}\,,\qquad \calk_{3,1}\ \to\ \calk_{3,1}+\beta\ \calk_{3,0}\,,
\eqlabel{zeromodes}
\end{equation}
are solutions as well. The constant $\beta$ is absorbed into the overall normalization of
the linearized fluctuations; and the constant $\alpha_1$ is fixed at order $\calo(b)$, similar to
how the constant $\alpha_0$ is fixed at order $\calo(\sqrt{b})$, as discussed below.

The asymptotic expansions near the boundary, \ie as $x\to 0_+$,
\begin{equation}
\begin{split}
&\calk_{3,0}= x^{3/4}\ \left(1+\frac{\sqrt{2}}{32} q_0\  x^{1/2}+\calo(x)\right)\,,
\end{split}
\eqlabel{br3uv1}
\end{equation}
\begin{equation}
\begin{split}
&F_{2}=x^{3/4} \left(f_{2;3,0}+\frac{\sqrt{2}}{96}\biggl(3\alpha_0 q_1  + q_0 (3 f_{2;3,0}-1)
\biggr)\ x^{1/2}+\calo(x)\right)\,,
\end{split}
\eqlabel{br3uv2}
\end{equation}
\begin{equation}
\begin{split}
&\calk_{3,1}=x^{3/4} \biggl(
\frac 32 \alpha_0 \ln x+k_{3,1;3,0}+x^{1/2} \left(\frac{3 \sqrt{2}\alpha_0q_0}{64} \ln x
-\frac{\sqrt{2}}{64}\left(3\alpha_0q_0-2k_{3,1,3,0}q_0-2q_1\right)  \right)
\\&+\calo(x\ln x)
\biggr)\,,
\end{split}
\eqlabel{br3uv4}
\end{equation}
\begin{equation}
\begin{split}
&\calk_{7,1}=x^{3/4} \biggl(-\alpha_0-\frac{ \sqrt{2}\alpha_0q_0}{128}\  x^{1/2}
+ \biggl(
k_{7,1;7,0}+\left(\frac{9\alpha_0}{20}-\frac{\alpha_0 q_0^2}{10240} \right) \ln x
\biggr)\ x\\
&+\calo(x^{3/2}\ln x) \biggr)\,,
\end{split}
\eqlabel{br3uv3}
\end{equation}
are characterized by
\begin{equation}
\left\{q_0\,,\, f_{2;3,0}\,,\, \alpha_0\,,\, q_1\,,\, k_{3,1;3,0}\,,\, k_{7,1;7,0}\right\}\,.
\eqlabel{b3para}
\end{equation}
As part of the overall normalization (choosing $\beta$ in \eqref{zeromodes}), we can set
\begin{equation}
k_{3,1;3,0}=0\,.
\eqlabel{fix2}
\end{equation}
Furthermore, $f_{2;3,0}=\alpha_1$ as defined in \eqref{zeromodes} --- it remains free at this order,
but is fixed at $\calo(b)$ order. 
Near the black hole horizon, \ie as $y\equiv 1-x\to 0_+$,
\begin{equation}
\begin{split}
&\calk_{3,0}=k_{3,0;0}^h+\calo(y^2)\,,\qquad F_{2}=f_{2;0}^h+\calo(y^2)\,,\\
&\calk_{3,1}=k_{3,1;0}^h+\calo(y^2)\,,\qquad
\calk_{7,1}=k_{7,1;0}^h+\calo(y^2)\,,
\end{split}
\eqlabel{br3ir}
\end{equation}
characterized by
\begin{equation}
\left\{k_{3,0;0}^h\,,\, f_{2;0}^h\,,\, k_{3,1;0}^h\,,\, k_{7,1;0}^h\right\}\,.
\eqlabel{b3parair}
\end{equation}
In total we have $6+4-1-1=8$ (from \eqref{b3para} and \eqref{b3parair} minus the constraint \eqref{fix2} and minus the free value of $\alpha_1$) parameters 
to numerically solve the system of four second-order ODEs \eqref{br31}-\eqref{br33}. 
The symmetry \eqref{zet2sym} implies the symmetry in $\calo(\sqrt{b})$ parameters:
\begin{equation}
\left\{\alpha_0,q_1,k_{7,1;7,0},f_{2;0}^h,k_{3,1;0}^h,k_{7,1;0}^h\right\}
\Longleftrightarrow \left\{-\alpha_0,-q_1,-k_{7,1;7,0},f_{2;0}^h,-k_{3,1;0}^h,-k_{7,1;0}^h\right\}\,.
\eqlabel{sympar}
\end{equation}
Discrete solutions are characterized by the number of nodes in the radial wavefunction $\calk_{3,0}$ ---
this is the tower of the excited QNMs. Additionally, $\zet_2$ symmetry \eqref{zet2sym}
implies that at order $\calo(\sqrt{b})$ there are actually 2 separate branches, $\calb_{3u}$
and $\calb_{3s}$, coalescing at $b=0$. This $\zet_2$ symmetry exchanges the branches, see \eqref{qdim3}. 
We focus on the lowest QNMs on each branch, \ie
we require that $\calk_{3,0}$ does not have nodes. For these lowest QNMs on $\calb_{3u}/\calb_{3s}$
(correspondingly $+/-$ signs in the table for $q_1$)
branches we find:
\begin{center}
\[
\begin{tabular}{|c|c|c|c|c|c|c|c|}
\hline
$q_0$  &  $\alpha_0$ & $q_1$  & $k_{7,1;7,0}$ & $k_{3,0;0}^h$ &  $f_{2;0}^h$ &  $k_{3,1;0}^h$ & $k_{7,1;0}^h$\\
\hline\hline
$-22.969$  &  $\mp 0.471 $ & $ \pm 27.434$  & $ \mp0.522$ & $0.350 $ &  $ -0.0432 $ &  $\pm 0.263 $ & $ \pm 0.261$\\
\hline
\end{tabular}
\]
\[
\]

\end{center}

Note that
\begin{equation}
\kk^2\equiv \frac{q}{(2\pi T)^2}=\frac{1}{16}\biggl(q_0\pm |q_1|\ \sqrt{b}\biggr)+\calo(b)\,.
\eqlabel{br3q2}
\end{equation}
The high-temperature approximations \eqref{br3q2} to the  QNM branches $\calb_{3u}/\calb_{3s}$ are
shown as dashed black/red lines on the right panel
of fig.~\ref{figure33ahigh}.

\bibliographystyle{JHEP}
\bibliography{ktqnm}

\end{document}